\documentclass[prl,aps,twocolumn,superscriptaddress,showpacs]{revtex4}
\usepackage{graphicx}
\def\(({\left(}
\def\)){\right)}
\def\[[{\left[}
\def\]]{\right]}

\begin{document}

\title{Temperature Chaos in Two-Dimensional  Ising Spin Glasses with Binary Couplings: a Further Case for Universality}

\author{Jovanka Lukic}
\affiliation{
Dipartimento di Fisica, 
Universit\`a di Roma ``Tor Vergata'', 
Via della Ricerca Scientifica,  
00133 Roma, Italy.}

\author{Enzo Marinari}
\affiliation{
Dipartimento di Fisica, INFM-CNR and INFN,
Universit\`a di Roma {\em La Sapienza}, 
P.le Aldo Moro 2, 
00185 Roma, Italy.}

\author{Olivier C. Martin}
\affiliation{
CNRS; Univ. Paris Sud, UMR8626, LPTMS, F-91405 Orsay CEDEX,  France.}

\author{Silvia Sabatini}
\affiliation{
Dipartimento di Fisica,
Universit\`a di Roma {\em La Sapienza}, 
P.le Aldo Moro 2, 
00185 Roma, Italy.}

\date{\today}

\begin{abstract}
\textbf{Abstract}  
We study temperature chaos in a two-dimensional Ising spin glass with
random quenched bimodal couplings, by an exact computation of the
partition functions on large systems.  We study two temperature
correlators from the total free energy and from the domain wall
free energy: in the second case we detect a chaotic
behavior. We determine and discuss the chaos exponent and the fractal
dimension of the domain walls.
\end{abstract}

\maketitle

%%%%%%%%%%%%%%%%%%%%%%%%%%%%%%%%%%%%%%%%%%%%%%%%%%%%%%%%%%%%%%%%%%%%%%%
\paragraph*{Introduction ---}
A characteristic feature of spin glasses is the presence of chaos
under small changes in the quenched couplings, in the temperature, or
in the magnetic field
\cite{berker,braymoore,banavarbray,kondor,niflehilhorst,nifleyoung,huse,bm,sasakimartin}.
With the expression \emph{temperature chaos} we refer to the
\emph{fragility} of the equilibrium states of a disordered system
under small temperature changes.  Let us consider two typical
equilibrium configurations of such a system under the same realization
of the quenched disorder: the first configuration is in equilibrium at
temperature $T$, while the second one is in equilibrium at temperature
$T' = T + \Delta T$.  One says that there is temperature chaos if for
arbitrarily small (but non-zero) values of $\Delta T$, the typical
overlap of two configurations at $T$ and $T'$ goes to zero when the
system size diverges. The spatial distance $\ell(T,\Delta T)$ over
which such overlaps decay is called the \textit{chaos length}, and, as
we will discuss better in the following, it scales as $\ell \sim
\Delta T^{-1/\zeta}$ when $\Delta T \to 0$.

The droplet picture \cite{braymoore,fisherhuse} of
spin glasses predicts that
\begin{equation}
\label{eq:zeta}
\zeta = \frac{d_s}{2} - \theta\;,
\end{equation}
where $d_s$ is the fractal dimension of the droplet interfaces and
$\theta$ is the usual spin glass stiffness exponent.  In
two-dimensional ($2D$) spin glasses with \emph{continuous}
distributions of spin-spin couplings, one has $d_s \approx 1.27$
(\cite{amorosomoore}) and $\theta \approx -0.285$, leading to the
prediction $\zeta \approx 0.92$; direct measurements of $\zeta$
\cite{braymoore,rieger} are in good agreement with this prediction.
The situation in the model with binary couplings ($J_{ij}= \pm 1$) is
notably different: no study of temperature chaos has been performed,
but if one works exactly at $T=0$ using the estimates
$d_s^{T=0}\approx 1.0$ (\cite{kardar}) and $\theta^{T=0}=0$
(\cite{hartmannyoung,amoruso}), one would expect $\zeta^{T=0} \approx
0.5$ which is very different from the continuous distribution case,
suggesting two universality classes. However, even though the $T=0$
properties of spin glasses with continuous and discrete distributions
are different, a new picture has recently emerged~\cite{joerg} in which
one shows that \emph{only one universality class} exists for $T>0$
critical properties when we consider the limit $T\to 0$.
In our work here we measure $\zeta^{T\to 0}$ in the $\pm J$ model and
show that its value is compatible with the one of the model with
Gaussian couplings, giving further credence to the generalized
universality scenario \cite{amoruso,joerg}.

Our study is based on exact partition function computations for large
but finite size systems, for many realizations of the quenched
disordered couplings \cite{galluccio1,galluccio2}.  Availability
of the full density of states of a large system is a very powerful
tool and it allows us to investigate the chaotic behavior of the free
energy for the $\pm J$ model. We also extract entropies of domain
walls and thus provide an improved estimate of the exponent $d_s$.

The outline of the paper is as follows. First we discuss the model,
our observables and the methods we use.  We discuss how to detect
temperature chaos.  We analyze what happens for the total free energy,
finding there is no temperature chaos there. Then we move on to domain
wall free energies. There we compute the two exponents, $d_s$ from the
entropy of $T=0$ domain walls, and $\zeta$ from $T>0$ scaling of two
temperature correlators. We close by concluding that $\zeta$ has a
value compatible with our generalized universality picture but not
with the naive application of Eq.~\ref{eq:zeta} (i.e. assuming
$T=0$ values for the domain-wall and chaos exponents).

%%%%%%%%%%%%%%%%%%%%%%%%%%%%%%%%%%%%%%%%%%%%%%%%%%%%%%%%%%%%%%%%%%%%%%%
\paragraph*{Model, observables and methods ---}
We analyze the $2D$ Edwards--Anderson spin glass with
Ising spins ($\sigma_i = \pm 1$). Its Hamiltonian has the form
\begin{equation}
H_J(\sigma) \equiv  - \sum_{<ij>} J_{ij}\, \sigma_i\, \sigma_j\;,
\label{eq:H1} 
\end{equation}
where the sum runs over all pairs of nearest neighbor sites on a
square $2D$ lattice of linear size $L$ with periodic boundary
conditions. The couplings $J_{ij}$ are independent quenched random
variables taking the two values $\pm 1$, each with probability $1/2$.

To analyze the possible chaotic behavior of the system, we look at
correlation functions of observables computed at two different
temperatures. For a generic observable ${\cal O}$ we use the
correlation function
\begin{equation}
\label{eq:correlation_funct2}
C_{{\cal O}}(L,T,\Delta T) \equiv
\frac{\overline{({\cal O}(T)- \overline{{\cal O}(T)})
                ({\cal O}(T')-\overline{{\cal O}(T')})} }
{\sqrt{\overline{({\cal O}(T)-\overline{{\cal O}(T)})^2}}
 \sqrt{\overline{({\cal O}(T')-\overline{{\cal O}(T')})^2}}}\;,
\end{equation}
where $T' \equiv T + \Delta T$, $L$ is the linear size of the lattice
and the bar stands for the quenched average over the distribution of
the couplings $J_{ij}$. In the rest of this note we will focus on two
main cases: in the first one ${\cal O}$ is the total
free energy $F$ of the system, while in the second case it is
the domain wall free energy, $F_{DW}$.

We consider a $L \times L$ lattice with given quenched couplings
$J_{ij}$, and we determine $F(T)$ and $F_{DW}(T)$ via the exact
computation of the partition function
$Z$~\cite{galluccio1,galluccio2,lukic}. We can then average over different
samples to obtain (up to some statistical errors) the correlation
functions $C_{{\cal O}} (L,T, \Delta T)$.

%%%%%%%%%%%%%%%%%%%%%%%%%%%%%%%%%%%%%%%%%%%%%%%%%%%%%%%%%%%%%%%%%%%
\paragraph*{The total free energy ---}

We start by discussing the computation of the free energy, where we
identify $\cal O$ with the total free energy $F$ of the system, and we
compute $C_F(L,T,\Delta T)$. In a system where the total free energy
has a chaotic behavior we would expect that in the large $L$ limit
$C_F$ should drop very fast~\cite{salesyoshino}, as a function of $L$,
even for infinitesimal values of $\Delta T$. In the following we
analyze $C_F$ and we do not find 
%that (somehow as expected, since $T_c=0$)
%one cannot detect 
a chaotic behavior.
% from comparing total free
%energies for different values of $T$. 
We use these results to show
that even in the $T\to 0$ limit the total free energy does not behave
chaotically in temperature.

We have averaged over a number of samples for different values of the
system size $V\equiv L\times L$. Specifically, we have collected
$99809$ samples at $L=10$, $31571$ samples at $L=20$, $3117$ samples
at $L=30$, $818$ samples at $L=40$ and $336$ samples at $L=50$.  Our
error bars have been computed by using the jack--knife method (for
details see \cite{fly}).

%As we have explained before i
In our approach we obtain the full density
of states, that allows us to compute expectation values at every
temperature (or couple of temperatures) we are interested in.  Because
of that we select our temperature ranges (both for $T$ and for
$\Delta T$) using the physical behavior of the system: we want
to be for example as much as possible in a scaling regime, i.e. at low
$T$, but not at very low $T$ \cite{lukic} (i.e. where, at given $L$,
the system already shows an unphysical behavior dictated by the
presence of a gap).  Because of these facts we will present in our
study data that span a range of temperatures going from $T = T_{min}
\simeq 0.2$ to $T = T_{max} \simeq 0.4$: these values are chosen such
that, for the considered linear sizes, we are in the scaling region
where the relevant physical phenomena occur.

\begin{figure}
\includegraphics[height=\columnwidth,angle=270]{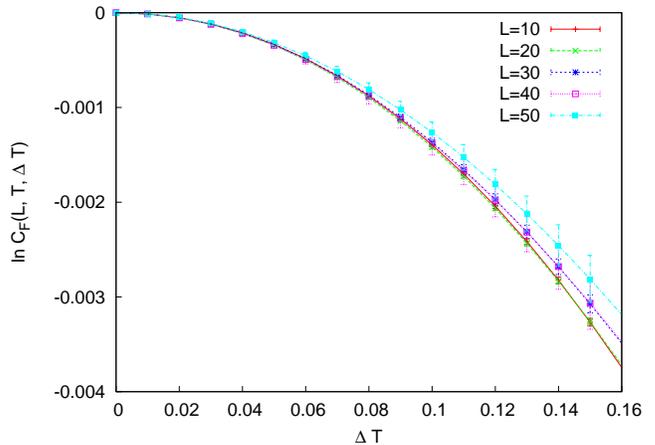}
\caption{$\log \{C_F(L,T,\Delta T)\}$ versus $\Delta T$, for
$T = 0.25$.\label{fig:uno}} 
\end{figure}

In figure \ref{fig:uno} we plot $\ln{C_F(L,T,\Delta T)}$ as a function
of $\Delta T$, for $T = 0.25$; (that we know from \cite{lukic,joerg} to
be in the scaling regime for $L\sim 50$); we use in the plot a $\Delta
T$ step of $0.01$ (reasonably small with respect to $T$).  Somehow
already this first, simple plot hints that the free energy of the
system does not unveil a chaotic behavior.  In fact one sees that the
correlation function \emph{rises} when $L$ is increased.  This trend
is even more evident for larger values of $\Delta T$.  Such a behavior
is the opposite of what we would expect in a chaotic scenario, where
for large $L$ values no correlation survives.

The issue of a possible presence of chaos in the free energy in the
limit $T\to 0$ (i.e. when $T$ approaches the $T=0$ critical point) has
to be dealt more carefully. For doing that we need a quantitative
analysis of our data.  We start by taking the infinite volume limit of
our correlation functions.  We present here the analysis of the
correlation function with $T=0.25$, but the other $T$ values follow
the same pattern.  We have analyzed a number of temperatures in the
scaling region that we are able to access (say for $T$ going from
$0.2$ up to $0.4$ \cite{lukic,joerg}).  For each value of $\Delta T$
we fit $C_F(L, T=0.25,\Delta T)$ to the functional form:
\begin{eqnarray}
\label{eq:correlation4} 
C_{F}(L,T = 0.25,\Delta T) & = & \widetilde{C}_F(T = 0.25, \Delta T) +  
\nonumber \\ 
                          & + & \frac{A(T=0.25,\Delta T)}{L}\;,
\end{eqnarray}
determining in this way the large $L$ limit $\widetilde{C}_F(T,\Delta T)$.
As we noticed already, the correlation function increases with $L$: the
fit to the form of Eq.~\ref{eq:correlation4} works very well, and $C_F$
reaches a strictly positive asymptotic value in the thermodynamic limit.

From there, we are able to study $\widetilde{C}_F(T,\Delta T)$ as a
function of $\Delta T$ (and, again, we do this for different values of
$T$). A very good fit is obtained by assuming the functional
dependence:
\begin{equation}
\label{eq:correlation5} 
\widetilde{C}_{F}(T,\Delta T) = \exp{(-a(T) \Delta T^2)}\;.
\end{equation}
In all cases this fit works very well; for example at $T=0.25$ we
estimate $a \simeq 0.114$.  We find that in the limit $T \to 0$ the
function $a(T)$ smoothly goes to a constant value (as opposed to its
possible divergence, that would indicate chaos in the free energy): in
the limit $T\to 0$ there is no chaos in the total free energy of the
system.

%%%%%%%%%%%%%%%%%%%%%%%%%%%%%%%%%%%%%%%%%%%%%%%%%%%%%%%%%%%%%%%%%%%%%%%
\paragraph*{Domain walls and their fractal dimension ---}
Our second observable is based on domain
walls.  We introduce a domain
wall into the system by applying to a given realization of the random
quenched couplings first periodic boundary conditions (pbc) and then
anti-periodic boundary conditions (apbc). For each sample, at any
temperature, we define the domain-wall free energy as the difference
of the free energies of these two systems; the same can be done for
the energy and entropy. So for each realization of the
couplings we compute the partition function twice: once with pbc and
once with apbc (equivalently, we compute the partition
function of two different systems with pbc, where in the second system
we have inverted the sign of one line of couplings of the first).

Before discussing our results for this interesting correlation
function we discuss a byproduct of this approach, that will be crucial
in our final discussion of the physics of two-dimensional spin
glasses.  A nice feature of the exact partition function approach is
that it gives us in particular the entropy of the domain wall even at
zero temperature: this is a quantity that vanishes in the models with
continuous $J_{ij}$ coupling values, but that is non-zero in the
model with binary quenched random couplings
where the ground state degeneracy is high.  

\begin{figure}
\includegraphics[height=\columnwidth,angle=270]{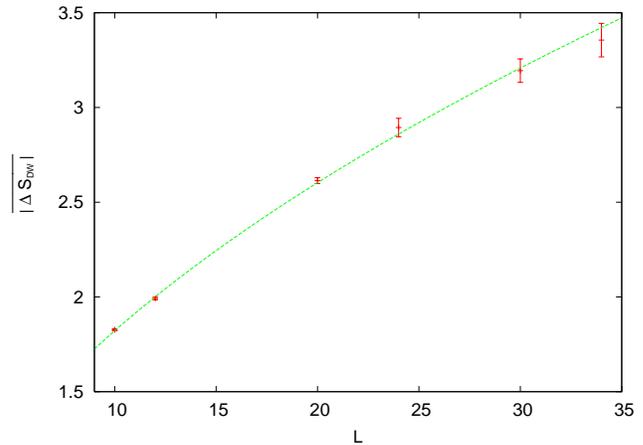}
\caption{$\overline{|\Delta S_{DW}|}$ versus $L$.} \label{fig:ds}
\end{figure}

The zero
temperature domain-wall entropy $\Delta S_{DW}$ has a disorder mean of
zero (since inverting a number of couplings maps us to another sample
in the disorder ensemble that has the same probability but has the
opposite value of $\Delta S_{DW}$).  We thus focus on the mean square
fluctuations; these have a characteristic size growing as
\begin{equation}
\overline{|\Delta S_{DW}|} \sim L^{\frac{d_s}{2}}
\end{equation}
when $L \to \infty$.  Previous work by Kardar and Saul~\cite{kardar}
(also based on partition function computations, but with smaller
statistics and for smaller $L$ values than us) gave the estimation
$d_s \approx 1.0$.  We have used our data to obtain a more precise
measurement of $d_s$. In Fig.~\ref{fig:ds} we show the scaling with
$L$ of the absolute value of $\Delta S_{DW}$ (to
exhibit the very accurate power scaling). We also show our best fit
which gives $d_s=1.03 \pm 0.02$ (where we only quote the statistical
error). This exponent $d_s$ is usually interpreted as the fractal
dimension of the domain wall created in the system; when $d_s>1$, the
domain wall is rough. However, in the $\pm J$ model there are many
ground states and so it is not appropriate to think of $d_s$ as
associated with one interface; in fact it is not a priori necessary
that $d_s \ge 1$.  We note again that for Gaussian quenched random
couplings one finds $d_s=1.27$ \cite{rieger,amorosomoore}; we thus see
that having many domain walls in our degenerate system leads to
smaller entropy fluctuations.

%%%%%%%%%%%%%%%%%%%%%%%%%%%%%%%%%%%%%%%%%%%%%%%%%%%%%%%%%%%%%%%%%%%%%%%%%
\paragraph*{The chaos exponent $\zeta$ ---}
Let us now discuss temperature chaos in the domain 
wall free energy $F_{DW}$, 
defined as the difference of the free energy in the system when using
periodic boundary conditions versus anti--periodic boundary
conditions. Given this quantity we use the definition
(\ref{eq:correlation_funct2}) and we compute $C_{F_{DW}}$. Notice that 
the relation (\ref{eq:correlation_funct2})
is simplified from the fact that
$\overline{F_{DW}}=0$.

\begin{figure}
\includegraphics[height=\columnwidth,angle=270]{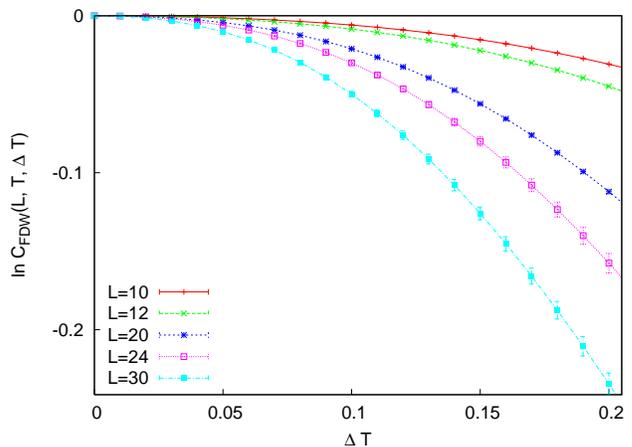}
\caption{$\log \{C_{F_{DW}}(L,T,\Delta T)\}$ versus $\Delta T$, for
$T = 0.25$.
\label{fig:tre}
} 
\end{figure}

Here we have used slightly smaller sizes than in the other case.  For
the linear sizes $L = 10$, $12$, $20$, $24$, $30$ we use different
numbers of samples: $99808$, $94351$, $31570$, $4098$, $3116$
respectively.  In figure \ref{fig:tre} we plot
$\ln{C_{F_{DW}}(L,T,\Delta T)}$ as a function of $\Delta T$ for
$T=0.25$.

Fig.~\ref{fig:tre} is very different from Fig.~\ref{fig:uno}: the
linear correlation coefficient of the domain wall free energies at two
different temperatures (with $\Delta T$ being fixed) decreases as $L$
increases.  As opposed to the case of the total free energy of the
system, this trend is compatible with an emergent chaotic behavior.
The same pattern arises when looking at the domain wall energy
($E_{DW} = E_{pbc} -E_{apbc}$) and at the domain wall
entropy, ($S_{DW} = S_{pbc} -S_{apbc}$) and analyzing
the related correlation functions $C_{E_{DW}}(L, T, \Delta T)$ and
$C_{S_{DW}}(L, T, \Delta T)$.

The fact that the total free energy does not show any sign of chaos
while the domain wall free energy hints for a possible chaotic
behavior is, as we have already discussed, very natural.  The small
value of the exponent $\theta$ implies that in the domain wall free
energy large cancellations play an important role \cite{braymoore}
\cite{banavarbray}. This suggests in turn that the domain wall energy and
the domain wall entropy (times $T$) exhibit a very similar temperature
dependence and greatly contribute to this cancellation.  This behavior
has as a natural consequence the presence of temperature chaos: as
soon as the parameters of the system get modified, even very
slightly, cancellations will act in a different way and select
completely different domain walls (the same arguments suggests
the presence of chaos under a small change in the quenched disorder).

In the scaling  theories, the domain wall free energy scales
with the lattice size as $L^{\theta}$ (where $\theta$ is the
stiffness exponent) and the domain wall entropy scales as
$L^{d_{s}/2}$ (where $d_{s}$ is the fractal surface of domain wall)
\cite{banavarbray}.  Consider first the system at temperature $T$. Here
the domain wall free energy scales with $L$ as:
\begin{equation}
\label{eq:free_energy} 
F_{DW}(T) \equiv E_{DW} - T S_{DW} 
\approx Y(T) L^{\theta}\;,
\end{equation}
where $Y(T)$ is the \emph{generalized stiffness coefficient}.
Now if the system is at temperature $(T + \Delta T)$, the
domain wall free energy scales as:
\begin{eqnarray}
\label{eq:free_energy2} 
F_{DW}(T + \Delta T) & = & E_{DW} - (T + \Delta T) S_{DW}\approx\nonumber\\
 & \approx & Y(T) L^{\theta} - \Delta T L^{d_{s}/2}\;.
\end{eqnarray}
Since the droplet theory predicts that $d_{s}/2 > \theta$, the
quantity (\ref{eq:free_energy2}) for large $L$ can change the sign
between $T$ and $T+\Delta T$. This implies that the equilibrium state
changes when one reaches a length $\ell$ such that
\begin{equation}
\label{eq:elle1} 
\ell^\theta \sim \Delta T\;\ell^{d_s/2}\;,
\end{equation}
that is
\begin{equation}
\label{eq:elle2} 
\ell\approx
\left(\frac{1}{\Delta T} \right)^{\frac1{d_{s}/2 - \theta}} \equiv
\left(\frac{1}{\Delta T} \right)^{\frac1{\zeta}}\;.
\end{equation}
It is important
to note that these scaling laws are valid for small $T>0$, and so
both $\theta$ and $d_s$ should be construed as being
obtained at $T>0$; this subtlety
will be essential for interpreting our data.

\begin{figure}
\includegraphics[height=\columnwidth,angle=270]{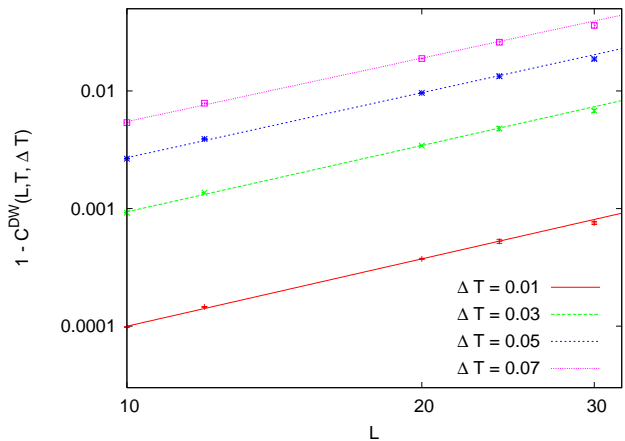}
\caption{${(1 - C_{F_{DW}})}$ vs $L$ in double logarithmic scale.
Here $T=0.25$.}
\label{fig:four}
\end{figure}

When $\Delta T$ is small and all relevant length scales diverge one
expects that \cite{niflehilhorst}
\begin{equation}
\label{eq:domain2} C_{F_{DW}}(L,T,\Delta T) \approx 1 -
\left(\frac{L}{\ell(T, \Delta T)} \right)^{2 \zeta}\;,
\end{equation}
for $L$ smaller than $\ell(T,\Delta T)$,
i.e. for small enough $\Delta T$.
In other words
in this limit $1-C_{F_{DW}}(L, T, \Delta T)$ scales 
as a power of $L$. In figure \ref{fig:four}
we plot ${(1 - C_{F_{DW}})}$ vs. $L$ on log-log scale
for small values of
$\Delta T$ ($\Delta T = 0.01$, $0.03$, $0.05$, $0.07$), 
when $T =0.25$
(similar results are obtained for other values of $T$). 
The scaling in $L$ allows us to obtain $\zeta$
and $\ell(T,\Delta T)$
from a linear fit to $\ln(1 - C_{F_{DW}})$; we obtain 
$\zeta$ values between
$0.92$ and $0.99$, depending on the values of
$T$ and $\Delta T$.

\begin{figure}[t]
\includegraphics[height=\columnwidth,angle=270]{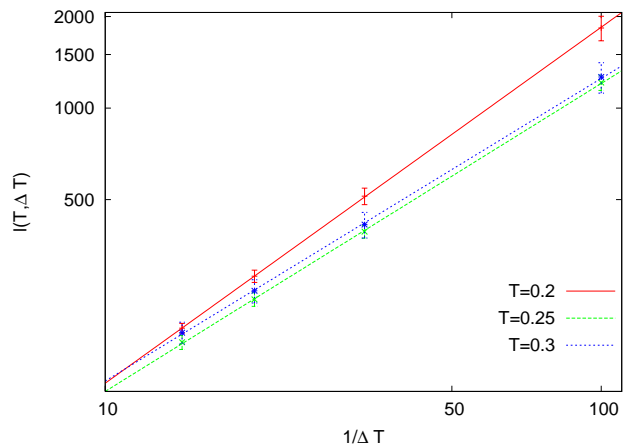}
\caption{$\ell(T,\Delta T)$ versus $\frac{1}{\Delta T}$ in double logarithmic scale.}
\label{fig:lDeltaT}
\end{figure}

We have also extracted from our fits the quantity $\ell(T,\Delta T)$.
Fig.~\ref{fig:lDeltaT}
shows that this length scale diverges as predicted in Eq.~\ref{eq:elle2}.
The different best fit values for $\zeta$ can be summarized by quoting 
$\zeta = 0.95\pm 0.05$; we do not detect any systematic 
dependence on $T$.

\begin{figure}
\includegraphics[height=\columnwidth,angle=270]{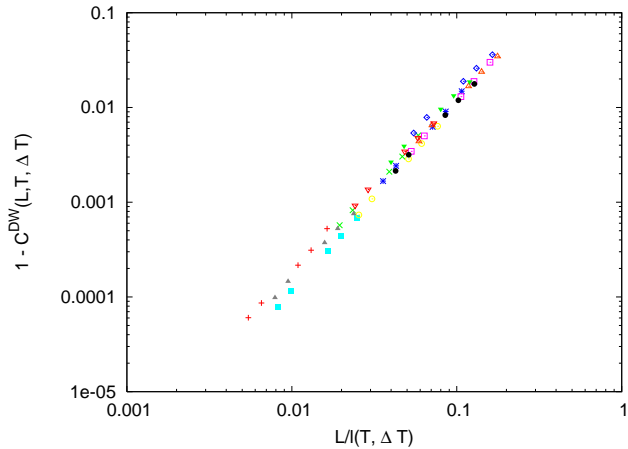}
\caption{$1-C_{F_{DW}}$ versus $L/\ell(T,\Delta T)$, for $T=0.2$, $T=0.25$ and
$T=0.3$ in double logarithmic scale.}
\label{fig:six}
\end{figure}

In Fig.~\ref{fig:six} we show an interesting data collapse.
We plot
$1-C_{F_{DW}}$ as a function of $L/\ell(T,\Delta T)$, for $T=0.25$ and
$T=0.3$ in double logarithmic scale. The fact that we get a single 
curve shows that the assumption that $C_{F_{DW}}(L, T, \Delta T)$ is
only a function of $L/\ell(T,\Delta T)$ is reasonable. The fact that the
curve is straight gives further support
to Eq.~\ref{eq:domain2}.

%

%%%%%%%%%%%%%%%%%%%%%%%%%%%%%%%%%%%%%%%%%%%%%%%%%%%%%%%%%%%%%%%%%%%%%%%%%5
\paragraph*{Discussion and conclusions ---}
This note spells a number of results. Let us start by summarizing
them.
First, when considering $2D$ Ising spin glasses with binary
couplings, the free energy $F$ does not show any temperature chaotic
behavior, even in the $T\to 0$ limit. This is in contrast~\cite{salesyoshino}
with what
happens in the directed polymer in a random medium, but in line with 
the prediction~\cite{sasakimartin} of the Migdal-Kadanoff approximation.
A simple interpretation is that the sample to sample non-chaotic
fluctuations of $F$ are $O(L)$, far larger than the chaotic
$O(L^{\theta})$ fluctuations arising from droplets.

As opposed to the total free energy,
the domain wall free energy does behave chaotically in
$T$. The question of interest is whether
the $\pm J$ model 
has chaotic behavior in the same class as the continuous
distribution models. We first measured the fractal dimension of domain walls
at $T=0$ exactly and found $d_s^{T=0}\sim 1.0$.
Then we measured directly the temperature 
chaos exponent $\zeta $ in the limit of small $T$
and found $\zeta=0.95\pm 0.05$.
(Note that these two measurements are conceptually very different.)

The point of view introduced by \cite{amoruso} and \cite{joerg} is the
following. If $T>0$, all $2D$ Ising spin glasses, independently of the
distributions of the quenched couplings, enjoy a strong, generalized
universality \cite{joerg}: a binary distribution of the couplings gives
rise to the same critical behavior as the Gaussian distribution.  Only
exactly at $T=0$ 
a very peculiar set of coupling distributions
(including binary couplings) generates an anomalous behavior
\cite{amoruso}: for cases where, by composing an arbitrary number of
couplings, an energy gap survives (for example a distribution where
$J=\pm 2, \pm 3$ is in this class, while a distribution where $J=\pm
\sqrt{2}, \pm \sqrt{3}$ is not), then at $T=0$ one gets $\theta=0$. Our
findings here for $T\to 0$ complete the scenario designed in
\cite{amoruso,joerg} and are completely consistent with it. If we use
indeed the relation implied by the droplet theory $\zeta = d_{s}/2 -
\theta$ and plug in both the value $\theta\sim -.285$ of the model
with Gaussian quenched couplings and the value $d_s\sim 1.27$ obtained
by \cite{hartmannyoung} and by \cite{amorosomoore} we would expect
$\zeta\sim 0.92$, in very good agreement with our estimate
$\zeta^{T\to 0}=0.95\pm 0.05$. We believe that this is a very
important point of support of the picture proposed and advocated in
\cite{amoruso}, \cite{joerg}: there is one single critical theory that
emerges, when $T\to 0$ in $2D$ spin glasses. The exponents
of this theory are, at this point, well determined.

Our measurements also confirm that $T=0$ is for the $J=\pm 1$
distribution, a singular point. In that model
at $T=0$, we found $d_s^{T=0}\sim 1.0$. This, 
together with the value $\theta = 0$,
implies that, exactly at $T=0$, if 
the relation $\zeta = d_{s}/2 - \theta$ were to be valid, we 
would have $\zeta^{T=0} = 1/2$. The
difference of the physics at $T=0$ and what one observes when
$T\to 0$ (as defined from the order used for taking the
$T\to 0$ and the $L\to\infty$ limits) is
at this point self-evident.

%%%%%%%%%%%%%%%%%%%%%%%%%%%%%%%%%%%%%%%%%%%%%%%%%%%%%%%%%%%%%%%%%%%%%%%%%5
\paragraph*{Acknowledgments ---}
This work was supported by the EEC's FP6
IST Programme under contract IST-001935, EVERGROW, and by the EEC's
HPP under contracts HPRN-CT-2002-00307 (DYGLAGEMEM) and
HPRN-CT-2002-00319 (STIPCO).

%\bibliographystyle{prsty}
%\bibliography{references}

\begin{thebibliography}{999}

\bibitem{berker}
McKay S.R., Berker A.N. and Kirkpatrick S.,
Phys. Rev. Lett. \textbf{48}, 767 (1982).

\bibitem{braymoore}
Bray A.J. and Moore M.A., 
Phys. Rev. Lett. \textbf{58}, 57 (1987).

\bibitem{banavarbray}
Banavar J.R. and  Bray A.J., 
Phys. Rev. B \textbf{35}, 8888 (1987).

\bibitem{kondor}
Kondor I., 
J. Phys. A \textbf{22}, L163 (1989).

\bibitem{niflehilhorst}
Ney--Nifle M. and Hilhorst H.J., 
Phys. Rev. Lett. \textbf{68}, 2992 (1992).

\bibitem{nifleyoung}
Ney--Nifle M. and Young A. P., 
J. Phys. A \textbf{30}, 5311 (1997).

\bibitem{huse}
Huse D.A. and Ko L.-F.,
Phys. Rev. B \textbf{56}, 14597 (1997).

\bibitem{bm}
Billoire A. and Marinari E., 
Europhys. Lett. \textbf{60}, 775 (2002), cond--mat/0202473.

\bibitem{sasakimartin}
Sasaki M. and Martin O.C., 
Phys. Rev. Lett. \textbf{91}, 245410 (2003), cond--mat/0310198.

\bibitem{fisherhuse}
Fisher D.S. and Huse D.A., 
Phys. Rev. B \textbf{38}, 386 (1988).

\bibitem{amorosomoore}
Amoruso C., Hartmann A.K., Hastings M.B. and Moore M.A., 
\textit{Conformal Invariance and SLE in Two-Dimensional Ising Spin Glasses},
preprint cond-mat/0601711.

\bibitem{rieger}
Rieger H., Santen L., Blasum U.. Diehl M., J{\"u}nger M. and Rinaldi G.,
J. Phys. A \textbf{29}, 3939 (1996).

\bibitem{kardar}
Saul L. and Kardar M.,
Nucl. Phys. B [FS] \textbf{431}, 641 (1994).

\bibitem{hartmannyoung}
Hartmann A.K. and Young A.P., 
Phys. Rev. B \textbf{64}, 180404 (2001).

\bibitem{amoruso}
Amoruso C., Marinari E., Martin O.C. and Pagnani A.,
Phys. Rev Lett. \textbf{91}, 087201 (2003), cond--mat/0305042.

\bibitem{joerg}
J\"org T., Lukic J., Marinari E. and Martin O.C.,
Phys. Rev Lett. \textbf{96}, 237205 (2006), cond-mat/0601480.

\bibitem{galluccio1}
Galluccio A., L\"obl M. and Vondr\'{a}k J., 
Phys. Rev. Lett. \textbf{84}, 5924 (2000).

\bibitem{galluccio2}
Galluccio A., L\"obl M. and Vondr\'{a}k J., 
Math. Program., Ser. A \textbf{90}, 273 (2001).

\bibitem{lukic}
Lukic J., Galluccio A., Marinari E., Martin O. C. and Rinaldi G.,
Phys. Rev Lett. \textbf{92}, 117202 (2004), cond--mat/0309238.

\bibitem{salesyoshino}
Sales M. and Yoshino H., 
Phys. Rev. E \textbf{65}, 066131 (2002).

\bibitem{fly}
Flyvbjerg H., {\sl Error Estimates on Averages of Correlated
Data}, in {\sl Advances in Computer Simulation}, edited by Kertesz
J. and Kondor I. (Springer, Berlin 1996).

\end{thebibliography}

\end{document}